\documentstyle[preprint,aps,epsf]{revtex}

\begin{document}

\tightenlines

\def\lqcd{\Lambda_{\rm QCD}}
\def\xslash#1{{\rlap{$#1$}/}}
\def\dsl{\,\raise.15ex\hbox{/}\mkern-13.5mu D}
\def\OMIT#1{}
\preprint{\vbox{\hbox{UCSD/PTH 97--19}
\hbox{UTPT-- 97-17}
\hbox{hep-ph/9708306}}}

\title{Renormalization Group Scaling of the $1/m^2$
HQET Lagrangian}

\author{Christian Bauer$^{a}$ and Aneesh V. Manohar$^{b}$}

\address{
\medskip (a) Department of Physics, University of Toronto\\
60 St.~George Street, Toronto, Ontario, Canada M5S 1A7\\
\medskip
(b) Department of Physics, University of California at San Diego,\\
9500 Gilman Drive, La Jolla, CA 92093-0319
\medskip
}
\bigskip
\date{August 1997}

\maketitle

\begin{abstract}
The operator mixing matrix for the dimension six operators in the heavy quark
effective theory Lagrangian is computed at one loop. The results are shown to
be consistent with constraints from the equations of motion, and from
reparametrization invariance.
\end{abstract}
\vskip2cm
Heavy quark effective theory (HQET) \cite{hqet1,hqet2} is a useful tool for
studying the physics of hadrons containing a single heavy quark. The HQET
Lagrangian has an expansion in powers of derivatives divided by the heavy quark
mass $m$, which translates into an expansion of hadronic quantities  in powers
of $\lqcd/m$, where $\lqcd$ is the non-perturbative scale parameter of the
strong interactions.

The HQET Lagrangian can be computed by matching with the full QCD Lagrangian at
a scale $\mu \approx m$. This has been done at one loop to order $1/m^3$
for the two-Fermion terms in the Lagrangian~\cite{eh,fgl,am}, and at one loop
to order $1/m^2$ for the two-Fermion terms~\cite{Blok}. The renormalization
group running of the dimension five ($1/m$) operators in the HQET Lagrangian
has been computed~\cite{eh,fgl}. There are several computations of the running
of the dimension six ($1/m^2$) operators
\cite{Blok,Lee,Finkemeier,Balzereit,Koerner} in the literature, but the various
papers disagree with each other. In
Refs.~\cite{Lee,Finkemeier,Balzereit,Koerner} the authors do not take into
account the effect of a four fermion operator which is present in the dimension
six operator basis and is related to the Darwin term by the equations of
motion. A complete calculation including all dimension six operators was given
in Ref.~\cite{Blok}, but we disagree with this calculation in a few entries of
the anomalous dimension matrix.

In this paper we compute the running of the dimension six operators of the HQET
Lagrangian. The coefficients are computed using one-loop running and tree-level
matching, which makes the calculation particularly simple. We will use the
notation of Ref.~\cite{Blok}, to make it easier to compare with previous
results. The computations are done in a theory with one heavy quark with
velocity $v$, $h_v$, and $n_f$ flavors of massless quarks, $q_i$. The covariant
derivative is chosen to be $\partial_\mu + i g A_\mu^A T^A$. With this
convention, the HQET Lagrangian to order $1/m^2$ is
\begin{eqnarray}
{\cal L}&=&-\frac{1}{4}G^{\mu\nu A}G_{\mu \nu}^A +
\sum_i \bar q_i i \dsl q + \bar{h}(iv\cdot D)h -\frac{c_k}{2m}\bar{h} D^2 h
-\frac{c_F}{4m}g\bar{h}\sigma_{\mu\nu}G^{\mu\nu}h \nonumber \\
&&+c_D O_D + c_S O_S +\sum c_i O_i,
\label{lagrangian}
\end{eqnarray}
where the Darwin and spin-orbit operators are defined as
\begin{equation}
 O_D = \frac{g}{8m^2}\bar{h}\left(D_\mu G^{\mu\nu}\right)v_\nu h,\qquad
 O_S = i \frac{g}{8m^2}\bar{h}\sigma_{\mu\nu}\{D^\mu,G^{\rho\nu}\}v_\rho,
\end{equation}
respectively. The remaining operators $O_i$ are four-Fermion operators
involving two heavy and two light-quark fields,
\begin{equation}
\begin{array}{rclrcl}
  O_1^{hl}&=&\displaystyle\frac{g^2}{8m^2}\sum_i\bar{h}
  T^A h\ \bar{q}_i\xslash v T^A q_i,\qquad &
  O_2^{hl}&=&\displaystyle\frac{g^2}{8m^2}\sum_i\bar{h}\gamma^\mu\gamma_5 T^A h
  \ \bar{q}_i\gamma_\mu\gamma_5 T^A q_i, \\
  \noalign{\medskip}
  O_3^{hl}&=&\displaystyle\frac{g^2}{8m^2}\sum_i\bar{h}
  h\ \bar{q}_i\xslash v q_i,&
  O_4^{hl}&=&\displaystyle\frac{g^2}{8m^2}\sum_i\bar{h}\gamma^\mu\gamma_5 h
  \ \bar{q}_i\gamma_\mu\gamma_5 q_i,\\
  \end{array}
\end{equation}
four-Fermion operators involving four light quark fields,
\begin{equation}
\begin{array}{rclrcl}
 O_1^{ll} &=&\displaystyle \frac{g^2}{8m^2}\sum_{i,j}\bar{q_i} T^A \gamma^\mu
 q_i \ \bar{q}_j T^A \gamma_\mu q_j,\qquad  &
 O_2^{ll} &=&\displaystyle \frac{g^2}{8m^2}\sum_{i,j}\bar{q_i} T^A \gamma^\mu
 \gamma_5 q_i \ \bar{q}_j T^A \gamma_\mu \gamma_5 q_j ,\\ \noalign{\medskip}
 O_3^{ll} &=&\displaystyle \frac{g^2}{8m^2}\sum_{i,j}\bar{q_i}  \gamma^\mu
 q_i \ \bar{q}_j \gamma_\mu q_j , &
 O_4^{ll} &=&\displaystyle \frac{g^2}{8m^2}\sum_{i,j}\bar{q_i} \gamma^\mu
\gamma_5
 q_i \ \bar{q}_j \gamma_\mu \gamma_5 q_j,\\
\end{array}
\end{equation}
the light quark penguin operator
\begin{equation}
O^p = \frac{1}{8m^2}\sum_i \bar q_i \gamma_\nu D_\mu G^{\mu \nu} q_i,
\end{equation}
and three-gluon operators
\begin{equation}
 O_1^{g} ={1\over 4 m^2}g f_{ABC}
 G_{\mu\nu}^A G^{B\mu}{}_\alpha G^{C\nu\alpha} ,\qquad
 O_2^{g} ={1\over 4 m^2} D^\mu\, G_{\mu \alpha}^A\, D_\nu\,  G^{\nu \alpha A}.
\end{equation}
The identity
\begin{equation}
 0= \int 2\ D^\mu\, G_{\mu \alpha}^A\, D_\nu\, G^{\nu \alpha A} +
 2\ g\, f_{ABC}\, G_{\mu\nu}^A G_{\mu\alpha}^B G_{\nu\alpha}^C +
 G_{\mu \nu}^A\, D^2\, G^{A\mu \nu}
\end{equation}
has been used to eliminate $G_{\mu \nu}^A\, D^2\, G^{A\mu \nu}$ from the
effective Lagrangian. There are also operators such as $\bar h \left( i v
\cdot D \right)^3 h$ which vanish by the heavy quark equations of motion, and
can be eliminated by a field redefinition of the Lagrangian. Operators $O^{hh}$
involving four heavy quark fields do not contribute in the single heavy quark
sector, and will be omitted from the Lagrangian in this analysis.

The equation of motion for the gluon field,
\begin{equation}
        D_\mu G^{\mu \nu A} = g v^\nu \bar h_v T^A h_v + g \sum_i
\bar{q_i} T^A \gamma^\nu q_i,
\end{equation}
can be used to further simplify the effective Lagrangian. It allows one to
rewrite $O_2^{g}$ and $O^p$ in terms of four-Fermion operators. It also gives
the
relation
\begin{equation}
O_D= O_1^{hl} +  \frac{g^2}{8m^2}\bar{h} T^a h\ \bar{h} T^a h.
\end{equation}
The second term on the right hand side can be omitted in the single heavy
quark sector, so the equation of motion reduces to
\begin{equation}\label{eom}
 O_D = O_1^{hl}.
\end{equation}
This relation can be used to further simplify the effective Lagrangian by
eliminating $O_1^{hl}$. For the moment, we retain $O_1^{hl}$, to make it easier
to
compare the results with Ref.~\cite{Blok}.

The renormalization group scaling of the $1/m^2$ terms in the HQET Lagrangian
involves the operators described above, as well as the time ordered product of
two $1/m$ operators.  The time ordered products will be denoted by
\begin{eqnarray}
O_{kk} &=& {i \over 2} \int d^4 x\ T\left[ O_k\left(x\right)\ O_k\left(0\right)
\right] ,\nonumber \\
O_{km} &=& {i} \int d^4 x\ T\left[ O_k\left(x\right)\ O_m\left(0\right)
\right] ,\nonumber \\
O_{mm} &=& {i \over 2} \int d^4 x\ T\left[ O_m\left(x\right)\ O_m\left(0\right)
\right] ,\label{top}
\end{eqnarray}
where the kinetic and magnetic moment operators are
\begin{equation}
O_k = -\frac{1}{2m}\bar{h} D^2 h ,\qquad
O_m = \frac{1}{4m}g\bar{h}\sigma_{\mu\nu}G^{\mu\nu}h.
\end{equation}
The coefficient of $O_k$, $c_k$, is fixed to unity by reparametrization
invariance~\cite{reparam}. The coefficient of the magnetic moment operator
satisfies the renormalization group equation~\cite{eh,fgl}
\begin{equation}\label{magrge}
\mu{d c_F \over d \mu} = 2 C_A c_F {g^2\over 16 \pi^2},
\end{equation}
where $C_A=3$ is the Casimir of the adjoint representation. We will later also
use
$C_F=4/3$, the Casimir of the fundamental representation (not to be confused
with
$c_F$), $T_F=1/2$, the index of the fundamental representation, $N=3$, the
number
of colors, and $n_f$, the number of light quark flavors. Note that the
coefficient of $O_m$ in the Lagrangian is $-c_F$.

The running of the dimension six operators is
\begin{equation}
 \mu {d  \over d\mu}
 \left(
 \begin{array}{c}
 O_{D} \\ O_S \\ O_{kk} \\ O_{km} \\ O_{mm} \\
 \tableline
 O^{hl}_1 \\ O^{hl}_2 \\ O^{hl}_3 \\ O^{hl}_4 \\
 \tableline
 O^{ll}_1 \\ O^{ll}_2 \\ O^{ll}_3 \\ O^{ll}_4 \\
 \tableline
 O_1^g \\
 \tableline
 O_{hh} \\
 \end{array}
 \right)= -{g^2\over 16\pi^2}\left(
 \begin{array}{cccccc}
 \Gamma_{11} & \Gamma_{12} & 0 & 0 & \Gamma_{15} \\
 \Gamma_{21} & \Gamma_{22} & 0 & 0 & \Gamma_{25} \\
 0 & \Gamma_{32} & \Gamma_{33} & 0 & \Gamma_{35} \\
 \Gamma_{41} & \Gamma_{42} & \Gamma_{43} & \Gamma_{44} & \Gamma_{45} \\
 0 & 0 & 0 & 0 & \Gamma_{55} \\
 \end{array} \right)
  \left(
 \begin{array}{c}
 O_{D} \\ O_S \\ O_{kk} \\ O_{km} \\ O_{mm} \\
 \tableline
 O^{hl}_1 \\ O^{hl}_2 \\ O^{hl}_3 \\ O^{hl}_4 \\
 \tableline
 O^{ll}_1 \\ O^{ll}_2 \\ O^{ll}_3 \\ O^{ll}_4 \\
 \tableline
 O_1^g \\
 \tableline
 O_{hh} \\
 \end{array}
 \right). \label{ans}
\end{equation}
The zero entries follow because no one-loop Feynman graph exists for that term
in
the mixing matrix. $O^{hh}$ represents all possible operators involving four
heavy
quark fields.

The tree-level matching conditions at the scale $\mu=m$ are $c_D=c_S=1$, and
that $c^{hl}_i$, $c^{ll}_i$ and $c^g_1$ are all zero. The operator
mixing equations Eq.~(\ref{ans}) then imply that $c^g_1$ and $c^{ll}_i$ stay
zero on scaling from $m$ to $\mu$. The values of $c_D$ and $c_S$ can be
obtained by solving the renormalization group equations in the $1-2$ sector.
The
equation of motion Eq.~(\ref{eom}) allows one to eliminate $O^{hl}_1$ and
further
simplify the calculation. The only $O^{hl}$ operator that can mix with $O_D$ or
$O_S$ is $O^{hl}_1$; there are no penguin diagrams from the other $O^{hl}_i$
operators. The $O^{hl}_i$ operators do not mix among themselves, so that
$\Gamma_{22}$ is diagonal~\cite{sumlogs}. Thus eliminating $O^{hl}_1$ using the
equations of motion gives a renormalization group equation for $c_D$ and $c_S$
only
in the $\left\{ O_{D}, \ O_S ,\ O_{kk} ,\ O_{km} ,\ O_{mm} \right\}$ sector.
The
computation of the coefficients is straightforward.
The renormalization group equation for the
spin-orbit coefficient $c_S$ is
\begin{equation}
 \dot c_S = {g^2\over 16 \pi^2} 4 C_A c_k c_F .
\label{csde}
\end{equation}
The running of $c_S$ is
consistent with the reparametrization constraint $c_S = 2 c_F-1$. The $c_S$
equation can be solved when combined with $c_k=1$ and Eq.~(\ref{magrge}) for
$c_F$~\cite{Blok,Lee,Finkemeier,Balzereit,Koerner},
\begin{equation}\label{cscfans}
c_S\left(\mu \right) = 2  z^{-C_A} - 1,\qquad c_F\left(\mu \right) = z^{-C_A},
\end{equation}
where
\begin{equation}
z = \left[ { \alpha_s\left(\mu \right) \over \alpha_s\left(m \right) }
\right]^{1/b_0},
\end{equation}
$b_0=11 C_A/3 -4 T_F n_f/3$ is the first term in the $\beta$-function, and
we have used the initial condition $c_S\left(m\right)=c_F\left(m\right)=1$.

The renormalization group equation for the Darwin term $c_D$ is
\begin{equation}
\dot c_D = {g^2\over 16 \pi^2}\left[\frac{13}{3} C_A c_D - \left( \frac{20}{3}
C_A +
\frac{32}{3} C_F \right) c_k^2 - \frac{1}{3} C_A c_F^2
 \right], \label{cdde2}
\end{equation}
whose solution is
\begin{equation}
 c_D\left(\mu \right) =
  z^{-2 C_A} +
 \left(\frac{20}{13} + \frac{32}{13}\frac{C_F}{C_A} \right)\left[1-z^{-13C_A/6}
 \right],
\label{soln}
\end{equation}
using the initial condition $c_D\left(m\right)=1$, and Eq.~(\ref{cscfans}).
For QCD, Eqs.~(\ref{cscfans},\ref{soln}) reduce to
\begin{equation}
c_S\left(\mu \right) = 2  z^{-3} - 1,\qquad c_F\left(\mu \right) =
z^{-3},\qquad c_D\left(\mu \right) =
z^{-6} +
\frac{308}{117}\left[1-z^{-13/2}  \right].
\end{equation}
As an example of numerical values,
running the
$b$-quark terms between $m_b$ and $m_c$ gives
\begin{equation}
c_F = 0.83,\qquad
c_S = 0.65,\qquad
c_D = 1.57,
\end{equation}
where we have used $n_f=4$ and $\alpha_s\left( m_c \right) / \alpha_s\left( m_b
\right) = 1.7$, to be compared with the tree-level values $c_F=c_S=c_D=1$.

To compute the full effective Lagrangian, including the four-Fermion operators,
one needs the anomalous dimension matrix Eq.~(\ref{ans}). The matrix is
computed without using the equations of motion to eliminate $O^{hl}_1$, to make
it easier to compare with earlier calculations. The entries of the anomalous
dimension matrix are listed below. The second form of the matrix uses the
explicit values of $C_A$, etc.\ for QCD. In deriving these equations, we have
used the identity
\begin{equation}
\bar \psi \left\{T^A,T^B\right\}\psi\ \bar \chi \left\{T^A,T^B\right\}
\chi = \left(1-1/N^2\right)\bar \psi \psi\ \bar \chi\chi + \left(N-4/N\right)
\bar \psi T^A \psi\ \bar \chi T^A \chi
\end{equation}
which is valid for Fermions in the fundamental representation of $SU(N)$, and
holds
regardless of the $\gamma$-matrix structure of the fermion bilinears.
\begin{equation}
\Gamma_{11} = \left(
\begin{array}{ccccc}
\frac{4}{3}C_A & 0 & 0 & 0 & 0 \\
0 & 0  & 0 & 0 & 0 \\
-\frac{2}{3}C_A-\frac{32}{3}C_F & 0 & 0 & 0 & 0 \\
0 & -4C_A & 0 & 2 C_A & 0 \\
-\frac{10}{3}C_A & 0 & 0 & 0 & 4 C_A \\
\end{array}\right) = \left(
\begin{array}{ccccc}
4 & 0 & 0 & 0 & 0 \\
0 & 0  & 0 & 0 & 0 \\
-{146}/{9} & 0 & 0 & 0 & 0 \\
0 & -12 & 0 & 6 & 0 \\
-10 & 0 & 0 & 0 & 12 \\
\end{array}\right),
\end{equation}
\begin{equation}
\Gamma_{12} = \left(
\begin{array}{cccc}
3C_A & 0 & 0 & 0 \\
0 & -3(N-4/N) &  0 & -3(1-1/N^2) \\
-6C_A & 0 & 0 & 0 \\
0 & -8(N-4/N) & 0 & -8 (1-1/N^2) \\
3C_A & -(N-4/N) & 0 & -(1-1/N^2)  \\
\end{array}\right) = \left(
\begin{array}{cccc}
9 & 0 & 0 & 0 \\
0 & -5 &  0 & -{8}/{3} \\
-18 & 0 & 0 & 0 \\
0 & -{40}/{3} & 0 & -{64}/{9} \\
9 & -{5}/{3} & 0 & -{8}/{9} \\
\end{array}\right),
\end{equation}
\begin{equation}
\Gamma_{21} = \left(
\begin{array}{ccccc}
{8}T_F n_f/{3} & 0 & 0 & 0 & 0 \\
0 & 0 & 0 & 0 & 0 \\
0 & 0 & 0 & 0 & 0 \\
0 & 0 & 0 & 0 & 0 \\
\end{array}\right)  = \left(
\begin{array}{ccccc}
{4}n_f/{3} & 0 & 0 & 0 & 0 \\
0 & 0 & 0 & 0 & 0 \\
0 & 0 & 0 & 0 & 0 \\
0 & 0 & 0 & 0 & 0 \\
\end{array}\right),
\end{equation}
\begin{equation}
\Gamma_{22} = \left(
\begin{array}{cccc}
-3 C_A & 0 & 0 & 0 \\
0 &-3 C_A & 0 & 0 \\
0 & 0 & 0 & 0 \\
0 & 0 & 0 & 0 \\
\end{array}\right) +2 b_0 = \left(
\begin{array}{cccc}
13 - \frac{4}{3}n_f & 0 & 0 & 0 \\
0 & 13 - \frac{4}{3}n_f & 0 & 0 \\
0 & 0 & 22 - \frac{4}{3}n_f & 0 \\
0 & 0 & 0 & 22 - \frac{4}{3}n_f \\
\end{array}\right),
\end{equation}
\begin{equation}
\Gamma_{32} = \left(
\begin{array}{cccc}
8C_F/3- 4 C_A/3 + 16 T_F n_f/3 & 0 & 0 & 0 \\
8 C_F/3 - 4 C_A/3 & 0 & 0 & 0 \\
{8}/{3} & 0 & 0 & 0 \\
{8}/{3} & 0 & 0 & 0 \\
\end{array}\right) = \left(
\begin{array}{cccc}
-{4}/{9}+{8}n_f/{3} & 0 & 0 & 0 \\
-{4}/{9} & 0 & 0 & 0 \\
{8}/{3} & 0 & 0 & 0 \\
{8}/{3} & 0 & 0 & 0 \\
\end{array}\right),\
\end{equation}
\begin{eqnarray}
\Gamma_{33} &=& \left(
\begin{array}{cccc}
\frac{8}{3}C_F-\frac{13}{3}C_A+\frac{16}{3}T_F n_f& 3(N-\frac{4}{N}) & 0
& 3(1-\frac{1}{N^2}) \\ \noalign{\smallskip}
\frac{8}{3}C_F-\frac{4}{3}C_A+3(N-\frac{4}{N}) & -3C_A &
3(1-\frac{1}{N^2}) & 0 \\ \noalign{\smallskip}
{8}/{3} & 12 & 0 & 0 \\
{44}/{3} & 0 & 0 & 0 \\
\end{array}\right)+2b_0 \nonumber \\ &=& \left(
\begin{array}{cccc}
{113}/{9}+{4}n_f/{3} & {5} & 0 & {8}/{3} \\
{41}/{9} & {13}-{4}n_f/{3} & {8}/{3} & 0 \\
{8}/{3} & 12 & 22 - {4}n_f/{3} & 0 \\
{44}/{3} & 0 & 0 & 22 - {4}n_f/{3} \\
\end{array}\right),
\end{eqnarray}
\begin{equation}\label{g44}
\Gamma_{41}=\Gamma_{42}=\Gamma_{43}=0,\qquad \Gamma_{44} = 12 C_A-2b_0 =
14 + {4}n_f/{3}.
\end{equation}
In writing Eq.~(\ref{ans}), penguin diagrams from $O^{ll}$ have been rewritten
in terms of $O^{ll}$ and $O^{hl}$ using the gluon equation of motion. The
Lagrangian
in the single heavy quark sector does not depend on $\Gamma_{i5}$, which have
not
been computed.

The anomalous dimension matrix Eq.~(\ref{ans}) has been computed previously.
$\Gamma_{11}$ was computed in
Refs.~\cite{Blok,Lee,Finkemeier,Balzereit,Koerner}. The submatrix in the $3-4$
sector was computed in Ref.~\cite{Blok,CS,jk,svz}. $\Gamma_{44}$ was computed
in Ref.~\cite{3gluon1,3gluon2}. The rest of the matrix was computed in
Ref.~\cite{Blok}. We disagree with the previous computations in a few entries
of the anomalous dimension matrix. One can check the consistency of
Eq.~(\ref{ans}) with the equation of motion Eq.~(\ref{eom}). One finds,
neglecting $O^{hh}$ operators which were also neglected in Eq.~(\ref{eom}),
\begin{eqnarray}
 \mu {d\over d\mu} \left(O_D - O^{hl}_1 \right) &=&
-{g^2\over16\pi^2} \left(\frac{4}{3}C_A O_D + 3 C_A O_1^{hl}\right) +
{g^2\over16\pi^2}
\left(\frac{8}{3} T_F n_f O_D + \left[ 2b_0-3 C_A \right] O_1^{hl}\right)
\nonumber
\\
 &=& -{g^2\over16\pi^2} \left( \frac{4}{3}C_A - \frac{8}{3}T_F n_f
\right)\left(O_D - O^{hl}_1
 \right),\label{check}
 \end{eqnarray}
so that the equation of motion is multiplicatively renormalized. This is
consistent with the result that equations of motion can only mix among
themselves under renormalization~\cite{politzer}. One can then use the
equations of
motion to eliminate $O_1^{hl}$. The resulting renormalization group equations
have the same form as Eq.~(\ref{ans}), with $O_1^{hl}$ omitted in the second
operator block. The new entries of the operator mixing matrix will be denoted
by $\Gamma_{ij}'$. The resulting matrix is (note that the zero entries are not
the same as Eq.~(\ref{ans}))
\begin{equation}
\left(
 \begin{array}{cccccc}
 \Gamma_{11}' & \Gamma_{12}' & 0 & 0 & \Gamma_{15} \\
 0 & \Gamma_{22}' & 0 & 0 & \Gamma_{25}' \\
 \Gamma_{31}' & 0 & \Gamma_{33} & 0 & \Gamma_{35} \\
 0 & 0 & 0 & \Gamma_{44} & \Gamma_{45} \\
 0 & 0 & 0 & 0 & \Gamma_{55} \\
 \end{array} \right),
\end{equation}
where the new entries are
\begin{equation}
\Gamma_{11}' = \left(
\begin{array}{ccccc}
\frac{13}{3}C_A & 0 & 0 & 0 & 0 \\
0 & 0  & 0 & 0 & 0 \\
-\frac{20}{3}C_A-\frac{32}{3}C_F & 0 & 0 & 0 & 0 \\
0 & -4C_A & 0 & 2 C_A & 0 \\
-\frac{1}{3}C_A & 0 & 0 & 0 & 4 C_A \\
\end{array}\right) = \left(
\begin{array}{ccccc}
13 & 0 & 0 & 0 & 0 \\
0 & 0  & 0 & 0 & 0 \\
-{308}/{9} & 0 & 0 & 0 & 0 \\
0 & -12 & 0 & 6 & 0 \\
-1 & 0 & 0 & 0 & 12 \\
\end{array}\right),
\end{equation}
\begin{equation}
\Gamma_{12}' = \left(
\begin{array}{ccc}
0 & 0 & 0 \\
-3(N-4/N) &  0 & -3(1-1/N^2) \\
0 & 0 & 0 \\
-8(N-4/N) & 0 & -8 (1-1/N^2) \\
-(N-4/N) & 0 & -(1-1/N^2)  \\
\end{array}\right) = \left(
\begin{array}{cccc}
0 & 0 & 0 \\
-5 &  0 & -{8}/{3} \\
0 & 0 & 0 \\
-{40}/{3} & 0 & -{64}/{9} \\
-{5}/{3} & 0 & -{8}/{9} \\
\end{array}\right),
\end{equation}
\begin{equation}
\Gamma_{22}' = \left(
\begin{array}{cccc}
-3C_A + 2b_0 & 0 & 0 \\
0 & 2b_0 & 0 \\
0 & 0 & 2b_0 \\
\end{array}\right) = \left(
\begin{array}{cccc}
13 - {4}n_f/{3} & 0 & 0 \\
0 & 22 - {4}n_f/{3} & 0 \\
0 & 0 & 22 - {4}n_f/{3} \\
\end{array}\right),
\end{equation}
\begin{equation}
\Gamma_{31}'= \left(
\begin{array}{ccccc}
8C_F/3- 4 C_A/3 + 16 T_F n_f/3 & 0 & 0 & 0 & 0\\
8 C_F/3 - 4 C_A/3 & 0 & 0 & 0 & 0\\
{8}/{3} & 0 & 0 & 0 & 0\\
{8}/{3} & 0 & 0 & 0 & 0\\
\end{array}\right) = \left(
\begin{array}{ccccc}
-{4}/{9}+{8}n_f/{3} & 0 & 0 & 0 & 0 \\
-{4}/{9} & 0 & 0 & 0 & 0 \\
{8}/{3} & 0 & 0 & 0 & 0 \\
{8}/{3} & 0 & 0 & 0 & 0 \\
\end{array}\right).\label{g13p}
\end{equation}

One can solve the renormalization group equations Eq.~(\ref{ans}) with the
intial conditions $c_F=c_D=c_S=1$, and $c^{hl}_i=c^{ll}_i=c^g_1=0$ to obtain
the coefficients in the effective Lagrangian with one-loop running, and
tree-level matching. The results are Eq.~(\ref{cscfans}) for $c_S$ and $c_F$,
and
\begin{eqnarray}
c_2^{hl} &=& \Bigl[N-\frac{4}{N}\Bigr]\Bigl[\frac{1}{2 b_0-7 C_A}
\left( z^{-2C_A}-z^{-b_0+3 C_A/2}\right) - \frac{2}{2 b_0-5 C_A}
\left( z^{-C_A}-z^{-b_0+3 C_A/2} \right) \nonumber\\
&& -\frac{3}{2 b_0-3 C_A} \left(1-z^{-b_0+3 C_A/2}
\right) \Bigr] \nonumber\\
c_3^{hl} &=& 0 \nonumber\\
c_4^{hl} &=& \left[1-\frac{1}{N^2}\right]
\left[\frac{1}{2 b_0-4 C_A} \left( z^{-2C_A}-z^{-b_0}
\right) - \frac{1}{b_0- C_A} \left( z^{-C_A}-z^{-b_0}
\right) -\frac{3}{2 b_0} \left(1-z^{-b_0}
\right) \right] \nonumber \\
c^{ll}_i &=& c^g_1 = 0.
\end{eqnarray}
One can solve Eq.~(\ref{ans}) for $c_D$ and $c^{hl}_1$ separately. The result
for the sum $c_D + c_1^{hl}$ is the same as the value of $c_D$ in
Eq.~(\ref{soln}), where the equation of motion was used to replace $O_1^{hl}$
by $O_D$. An independent linear combination of $c_D$ and $c^{hl}_1$
that has a relatively simple scaling law is
\begin{eqnarray}
c_D - \frac{8T_F n_f}{9 C_A} c_1^{hl} &=& -\frac{5 C_A + 4 T_F n_f}{4 C_A + 4
T_f n_f} z^{-2 C_A} +
\frac{C_A +16 C_F - 8 T_F n_f}{2(C_A-2T_F n_f)}\nonumber \\
&&+ \frac{-7 C_A^2 + 32 C_A C_F - 4 C_A T_F n_f +32 C_F T_F
n_f}{4(C_A + T_F n_f)(2 T_F n_f-C_A)} z^{4 T_F n_f/3 - 2c_A/3} .
\end{eqnarray}
The numerical values for the coefficients with the same choice of parameters
as before are $c_2^{hl}=-0.19$, $c^{hl}_3=0$, $c^{hl}_4=-0.09$, $c_D = 1.61$
and $c^{hl}_1=-0.04$.

We would like to thank D.~Pirjol for several helpful email messages about the
anomalous dimension matrix, and for rechecking some of the results of
Ref.~\cite{Blok}. C. B. is grateful to M.E.~Luke for numerous discussions on
this topic. This work was supported in part by a Department of Energy grant
DOE-FG03-97ER40546.

\end{document}